\documentclass[12pt]{iopart}
\usepackage{epsfig,cite}
\begin{document}

\title[Phase-imprinted BEC tunnelling in time-dependent double-well potential]{Josephson tunnelling of a phase-imprinted Bose-Einstein condensate in a time-dependent double-well potential}

\author{E. Sakellari\footnote[3]{To whom correspondence should be addressed (Eleni.Sakellari@durham.ac.uk)},
N. P. Proukakis, M. Leadbeater, and C. S. Adams}

\address{Department of Physics, University of Durham, Durham DH1 3LE, United Kingdom}

\begin{abstract}

This paper discusses the feasibility of experimental control of the flow direction of atomic Bose-Einstein condensates 
in a double-well potential using phase-imprinting. The flow is induced by the application of a 
time-dependent potential gradient, providing a clear signature of macroscopic quantum tunnelling in atomic condensates. 
By studying both initial state preparation and subsequent tunnelling dynamics we find the parameters to optimise the phase 
induced Josephson current. We find that the effect is largest for condensates of up to a few thousand atoms, 
and is only weakly-dependent on trap geometry.

\end{abstract}


\maketitle

\section{Introduction}

The creation of superconducting \cite{SC_WeakLink} and superfluid \cite{SF_WeakLink} weak links has led to the 
experimental observation of Josephson effects \cite{Josephson}, arising as a result of macroscopic quantum phase coherence. 
Josephson weak links are typically created by connecting two initially independent 
superconducting or superfluid systems via a barrier with dimensions of the order of the system healing length. 
Such junctions lead to a variety of interesting phenomena \cite{barone}, including dc- and ac-Josephson effects. 
Observations in superconductors preceded those in superfluids, due to the much larger 
healing lengths, thus enabling easier fabrication of weak links. Evidence for Josephson-like effects has been observed in 
$^{4}$He weak links \cite{Sukhatme}, and unequivocally demonstrated for weakly-coupled $^{3}$He systems \cite{avenel}. 
The recent achievement of dilute trapped atomic Bose-Einstein condensation (BEC) 
\cite{BEC}  gives rise to a new system for studying Josephson effects. In particular, such systems enable the investigation of
 dynamical regimes not easily accessible with other superconducting or superfluid systems. 
Remarkable experimental progress has led to the creation of atomic BEC Josephson junction arrays, in which 
the harmonically trapped atoms are additionally confined by an optical lattice potential, 
generated by far-detuned laser beams. Phase coherence in different wells was observed by 
interference experiments of condensates released from the lattice \cite{Kasevich}. 
In addition, Josephson effects \cite{Inguscio} and the control of tunnelling rate have been 
demonstrated \cite{Mott,NIST}. Alternative insight into the diverse range of Josephson phenomena 
can be obtained by looking at a single Jospehson junction arising in a double-well system. 
This system has already received considerable theoretical attention, with treatments based on a two-state approximation 
\cite{Two_State_0,Two_State_1,Two_State_2,shen,Pendulum,Pi_BEC,Two_State_3,Two_State_4,Two_State_5,Janne,kohler,Sols_NJ}, 
zero temperature mean field theory \cite{MF_1,MF_2,leggett1,garraway,fant,Two_State_6,MF_3,leggett2,sak},
 quantum phase models \cite{Phase,Exact_Phase} and instanton methods \cite{Instanton}. 
Experimentally, a double well potential may be produced by adding a blue-detuned laser beam which induces 
a repulsive gaussian barrier to a harmonic trap \cite{MIT_DW}. 
Atomic interferometry based on such a set-up was recently reported \cite{MIT_Interferometer,mit_interferometer_2}. 
Alternatively, a condensate can be created directly in a magnetic double-well structure \cite{foot,Walraven}.

In this paper, we investigate the Josephson dynamics for a phase-imprinted atomic 
condensate in a double-well potential under the influence of a time-dependent potential gradient. 
We focus on the sensitivity of the Josephson flow direction to the initial state preparation. In particular, preparation 
in the odd parity energy eigenstate combined with the application of the potential gradient leads to a flow towards a 
region of higher potential energy providing a clear signature of Josephson tunnelling. Flow 
reversal in context of the Josephson effect is well known. For example in a 
superconducting $\pi$-junction \cite{Pi_Junction}, the addition of a macroscopic phase difference 
$\phi=\pi$ across the superconducting weak link leads to reversal of the sign of the current \cite{Reversal,SQUID,Fluxons}. 
Also a related effect has been predicted for condensates in optical lattices 
as a result of the renormalization of the mass in the lattice, based on Bloch wave analysis \cite{levit}.

The superfluid analogue of a supeconducting $\pi$-junction is a metastable 
$\pi$-state, recently observed in $^{3}$He weak links \cite{Pi_State}. 
Atomic BEC junctions behave similarly to those of $^{3}$He-B. Thus, although
 superconducting Josephson junctions can be mapped onto a rigid pendulum, atomic condensate tunnel 
junctions map onto a non-rigid pendulum \cite{Pendulum,Two_State_2,shen}, thus exhibiting richer oscillation modes. 
For example, $\pi$ oscillations arise in such systems 
\cite{Pendulum,Two_State_2,shen,Pi_BEC,Exact_Phase}. For an atomic condensate in a double-well potential, 
$\pi$ oscillation modes can be produced by imprinting a phase shift of $\pi$ between the two wells. We study 
how these modes behave under the action of an external potential difference. 

The paper is structured as follows. Sec. 2 introduces our main formalism, 
and briefly reviews the low-lying states of a condensate 
in a double-well, while Sec. 3 discusses the dynamics associated with particular initial
states, including the $\pi$ oscillation modes, under the addition of a time-dependent potential gradient. 
The experimental observation of controlling the Josephson flow direction by 
phase imprinting in current BEC set-ups is analysed in Sec. 4, with a short conclusion in Sec. 5.

\section{Time-independent Properties of a BEC in a double-well potential}

At low temperatures, the behaviour of a Bose-Einstein condensate is accurately described by a 
nonlinear Schr\"{o}dinger equation known as the Gross-Pitaevskii (GP) equation. Throughout this 
paper we work in dimensionless (harmonic oscillator) units, by applying the following scalings:space coordinates
 transform according to 
$\mbox{\boldmath$r$}_{i}^{\prime}={a_{\perp}}^{-1}\mbox{\boldmath$r$}_{_{i}}$, time $t^{\prime}=\omega_{\perp} t$, 
condensate wavefunction 
$\psi^{\prime}\left(\mbox{\boldmath$r$}^{\prime},t^{\prime}\right)=\sqrt{a_{\perp}^3}\psi\left( \mbox{\boldmath$r$},t\right)$
  and energy $E^{\prime}=\left(\hbar\omega_{\perp}\right)^{-1}E$. 
Here $a_{\perp}=\sqrt{\hbar/m\omega_{\perp}}$ is the harmonic oscillator length in the transverse direction(s),
 where $\omega_{\perp}$ the corresponding trapping frequency. 
We thus obtain the following dimensionless GP equation (primes henceforth neglected for convenience) 
describing the evolution of the condensate wavefunction (normalised to unity)
\begin{eqnarray}
i \partial_{t}\psi \left( \mbox{\boldmath$r$},t\right) =  \left[ -{\frac{1}{2}} \nabla^{2}
+V\left( \mbox{\boldmath$r$}\right)+\tilde{g}|\psi(\mbox{\boldmath$r$},t)|^{2}\right] \psi (\mbox{\boldmath$r$},t)~.
\label{eq:GP3dhou}
\end{eqnarray}
The atom-atom interaction is parametrized by $\tilde{g}=g/(a_{\perp}^3 \hbar \omega_\perp)$, where
 $g={\cal N}(4\pi\hbar^2a/m)$ is the usual three-dimensional scattering amplitude, 
defined in terms of the {\it s}-wave scattering length $a$, and ${\cal N}$ is the total number of atoms (mass $m$).
The total confining potential (see Fig. 1(a)) is given by 
\begin{equation}
V\left({\bf r}\right)=\textstyle{1\over 2}\left[(x^2+y^2)+\lambda^2 z^2\right]+%
h\exp\left[-(z/w)^2\right]+\delta z~.
\label{eq:conf3d}
\end{equation}
The first term describes a cylindrically symmetric harmonic trapping potential, with a trap aspect ratio 
$\lambda=\omega_{z}/\omega_{\perp}$. The trap is spherical for $\lambda=1$, 
`cigar-shaped' for $\lambda<1$  and `pancake-like' for $\lambda>1$. The second term describes a gaussian potential of height $h$ generated by a blue detuned light sheet of beam waist $w$ in the $z$ direction,
 located at $z=0$. In Eq. (2), the contribution $\delta z$ corresponds to an additional linear potential of 
gradient $\delta$ pivoted at the centre of the trap. For $\delta>0$, the right well obtains higher potential energy 
and the trap centre is additionally shifted into the $z>0$ region; however, this shift is negligible
 for the parameters studied throughout this work, and will be henceforth ignored.

\begin{figure}[hbt]
\centering
\includegraphics[width=12.0cm]{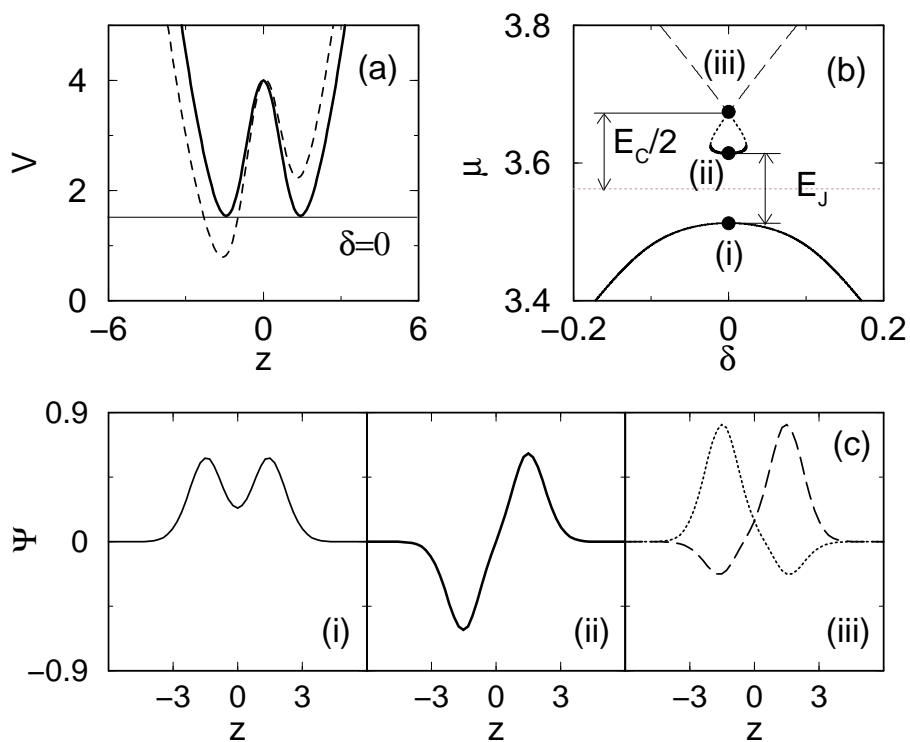}
\caption{ Double well potential with corresponding eigenenergies
and eigenstates. (a) Schematic geometry of the total  confining 
potential in the axial direction for a Gaussian barrier (height $h=4 \hbar \omega_{\perp}$, waist $w=a_{\perp}$) located at the
centre of the trap. Plotted are the symmetric  ($\delta=0$, solid line) and asymmetric ($\delta =0.5 (\hbar \omega_{\perp} /
a_{\perp})$, dashed line) cases. (b) Eigenenergies for the double-well as a function of the 
potential gradient $\delta$ indicating the self-interaction energy, $E_{\rm C}$, and the Josephson
 coupling energy, $E_{\rm J}$. The horizontal dotted grey line corresponds to the zero energy of the two-state model. 
The parameters used here are $\tilde{g}=\pi$ and spherical trap 
geometry ($\lambda=1$) corresponding to $E_{\rm C}=0.220 \hbar\omega_\perp $ and $E_{\rm J}=0.102 \hbar\omega_\perp$. 
(c) Eigenstates at the centre of the trap: (i) ground state (lower solid 
line), (ii) anti-symmetric first-excited state with equal population in both wells (thick solid 
line), (iii) first excited state with unequal populations, having 
more population in left well (dotted), or in right well (dashed).}
\end{figure}

The eigenestates of the double-well condensate are calculated by substituting $\psi\left(\mbox{\boldmath$r$},t\right)=
e^{-i\mu t}\phi \left(\mbox{\boldmath$r$}\right)$ and solving the resulting time-independent equation as discussed in 
\cite{sak}. As is well-known, sufficiently large interactions lead 
to the appearance of a loop structure, see e.g. \cite{Two_State_5,Loop}. The 
loop structure for the first excited state is shown in Fig. 1(b). 
Corresponding wavefunctions for ground and first excited state are 
shown in Fig. 1(c) for $\delta=0$. The three eigenstates are (i) 
asymmetric ground state $\Psi_{g}$ with equal population in both wells, (ii) an 
anti-symmetric state with equal population in both wells and a phase difference of $\pi$ across the trap centre, 
which we shall henceforth refer to as  $\Psi_{e}$ and (iii) two higher energy `self-trapped' states with most of the 
population in either the left (dotted) or the right (dashed) well \cite{Two_State_1,Two_State_2,shen,sak}. 
This paper is mainly concerned with states $\Psi_{g}$ and $\Psi_{e}$ and superpositions thereof. 
In particular, we will show that the dynamics of excited states in the presence 
of a time-dependent potential $\delta z$ are remarkably different to that of the ground
 state, and offer a clear demonstration of Josephson tunnelling.

\section{Tunnelling Dynamics under a time-dependent magnetic field gradient}

The main theme of the present study is to consider the tunnelling of states with initial phase $\phi=0$ and $\pi$, whose
symmetry is broken by the addition of a time-dependent 
potential gradient which increases linearly. 
The potential gradient is applied at $t=0$, i.e., $\delta=Rt$ for $t>0$, 
(dashed line in Fig.1(a) showing the right well with higher potential energy than the left). 

Before considering the effect of asymmetry, we review
the behaviour of the symmetric double-well.
The solution of the GP equation in a double-well potential can be mapped onto 
a two-state model 
\cite{Two_State_0,Two_State_1,Two_State_2,shen,Pendulum,Pi_BEC,Two_State_3,Two_State_4,Two_State_5,Janne,kohler,Sols_NJ,
MF_1,MF_2,leggett2,garraway,MF_3,leggett1,fant,Two_State_6,sak} 
by writing the 
wavefunction as a superposition of states localised in the left and
right wells, $\psi_\ell(\mbox{\boldmath$r$})$ and $\psi_r(\mbox{\boldmath$r$})$,
i.e.,
\begin{equation}
\Psi\left( \mbox{\boldmath$r$},t\right)=
c_\ell(t)\psi_\ell(\mbox{\boldmath$r$})+c_r(t)\psi_r(\mbox{\boldmath$r$})~. 
\label{eq:tsmansanz}
\end{equation}
The Hamiltonian for this two-state system is,
\begin{equation}
H=\frac{1}{2}\left[\begin{array}{cc}
-\Delta+E_{\rm C}N & -E_{\rm J} \\
-E_{\rm J}        & \Delta-E_{\rm C}N
\end{array}\right]~,
\label{eq:hamiltonia}
\end{equation}
where $N= (N_\ell-N_r) /{\cal N}$ is the fractional population difference between the left and right well, 
$\Delta$ is the potential energy difference between the left and right well ($\Delta=\alpha\delta$, where
 $\alpha$ is a numerical factor determined numerically from the GP solution), 
$E_{\rm C}=\tilde{g}\langle\psi_{\ell,r} | |\psi_{\ell,r}|^{2}|\psi_{\ell,r}\rangle$ is the self-interaction energy, 
$E_{\rm J}=-2\langle\psi_\ell\vert\left(-\textstyle{1\over2} \nabla ^2+V_{\delta=0}\right)\vert\psi_r\rangle$ 
is the Josephson coupling energy. The energy splittings $E_{\rm C}$ and $E_{\rm J}$ are indicated in Fig. 1(b). 
By writing $c_i=\sqrt{N_i}\exp(i\phi_i)$ and defining a relative phase $\phi=\phi_\ell-\phi_r$ 
the two-state Schr\"odinger equation can be re-written in terms of the coupled equations 
\begin{equation}
\dot{N}=E_{\rm J}\sqrt{1-N^2}\sin\phi~,~{\rm and}~
\dot{\phi}=\Delta-NE_{\rm C}-\frac{E_{\rm J}N}{\sqrt{1-N^2}}\cos\phi~.
\label{eq:2state}
\end{equation}
To find the dynamics of $N$ and $\phi$ one needs to know the value of $E_{\rm C}$, $E_{\rm J}$ and $\Delta$ for any particular 
barrier height, asymmetry and nonlinearity. Below (Fig. 3) we confirm that the two-state model is an excellent
 approximation to the full solution of the Schr\"odinger equation as long as $\vert \delta \vert$ is not too large.

If a system is initially prepared in one of its eigenstates, $\Psi_g$ or $\Psi_e$, it will remain in that same state 
and there is no tunnelling current. This is shown in Fig. 2(a), where we plot the fractional relative 
population, $N$ as a function of time with $\delta=0$ for $t<0$. However, if the system is prepared in 
a superposition of $\Psi_g$ or $\Psi_e$ with a $\pi$ phase difference, i.e.,
\begin{equation}
\Psi_{\pi\pm}=\frac{1}{\sqrt2}(\Psi_g \pm \Psi_e)~,
\end{equation}
the population tunnels back and forth (see Fig. 2(b) and (c)) and the relative phase between the two wells 
oscillates around a mean value of $\pi$ ($\pi$-oscillations \cite{Pendulum,Two_State_2,shen,Pi_BEC,Exact_Phase}). 
The amplitude of the $\pi$-oscillations depends on the ratio $\Lambda=E_{\rm C}/E_{\rm J}$
 \cite{shen,Pendulum,Pi_BEC,leggett1,MF_3,leggett2}. By solving the two-state coupled equations (\ref{eq:2state}) 
with initial conditions $N(0)=0.994$ (determined from the GP 
solution for $\psi_{\pi-}$) and $\phi(0)=\pi$, we find a critical ratio, 
$\Lambda_c \sim 1.80$. For $\Lambda<\Lambda_c$, the population oscillates between 
$\pm N(0)$, as in Fig.~2(b)($t<0$), whereas for $\Lambda>\Lambda_c$ the oscillations in $N$ are suppressed Fig.~2(c)($t<0$). 
Note that in Fig.~2(b) we have shifted the time origin by a quarter of an oscillation period such that $N=0$ at $t=0$. 

The effect of introducing an asymmetry depends sensitively on the initial 
state. For $\Psi_g$, the potential gradient induces a Josephson current to the left (lower potential energy region), 
whereas for $\Psi_e$ flow occurs to the right (higher potential energy) (Fig. 2(a)). 
The situation is more complex for superposition states, such as $\Psi_{\pi\pm}$. For $\Lambda<\Lambda_c$, if 
the potential gradient is turned on rapidly (compared to the period of the $\pi$-oscillations), 
the oscillations in $N$ are suppressed tending towards a mean $N$ close to its initial value, Fig. 2(b). 
For $\Lambda>\Lambda_c$, the induced flow is very different for $\Psi_{\pi +}$ and $\Psi_{\pi -}$. For $\Psi_{\pi +}$ most 
of the population remains self-trapped in the higher energy well, 
whereas for $\Psi_{\pi -}$ a large fraction of the population flows from the lower to the upper well. 

The parameters used throughout the rest of the paper, $\tilde{g}=\pi$ and a spherical trap 
geometry ($\lambda=1$), give $E_{\rm C}=0.220 \hbar\omega_\perp $ and 
$E_{\rm J}=0.102 \hbar\omega_\perp$, corresponding to the regime, $\Lambda>\Lambda_c$. 
In this regime, the experimental preparation of $\Psi_{\pi \pm}$ 
from the ground state is more difficult as it requires a transfer 
of population in addition to imprinting a phase diference of $\pi$. However, as we will see in Sec. 4,
 straightforward phase-imprinting can generate a superposition that contains a large fraction of $\Psi_{\pi \pm}$.

\begin{figure}[t]
\centering
\includegraphics[width=12.0cm]{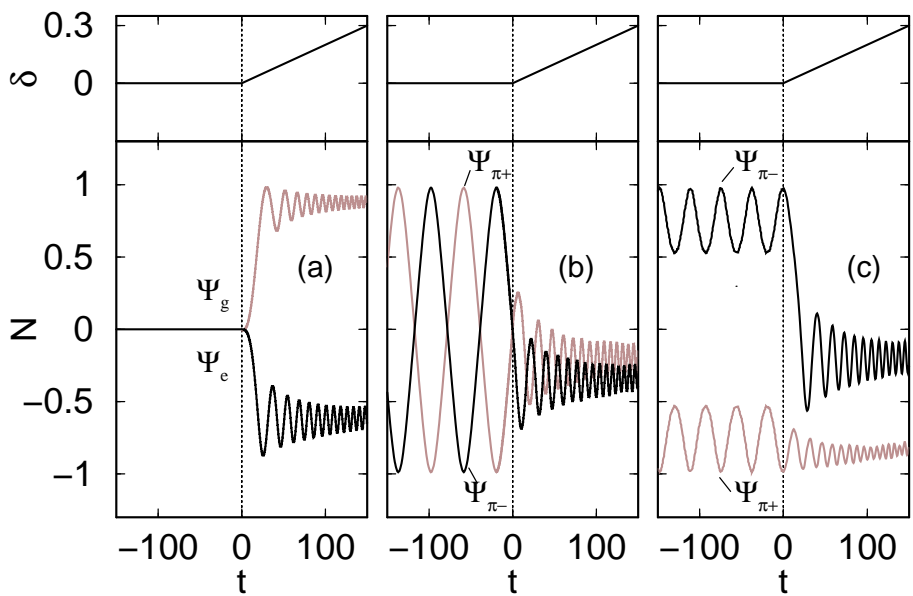}
\caption{Evolution of fractional population difference $N$ as a function of time
(calculated using the GP equation) without ($t<0$) and with ($t>0$) a potential
gradient $\delta=Rt$  (shown at the top of each figure)
for different initial states: (a) a pure ground $\Psi_g$ or first
excited $\Psi_e$ state with equal populations in both wells. 
In this case, tunnelling only arises due to the additional
external potential; (b) the superposition states $\Psi_{\pi\pm}$ for $\tilde{g}=\pi/2$,
corresponding to the regime $\Lambda<\Lambda_c$ 
($E_{\rm C}=0.123 \hbar\omega_\perp $ and $E_{\rm J}=0.095 \hbar\omega_\perp$),
showing maximum amplitude $\pi$-oscillations for $t<0$:
and (c) the superposition states $\Psi_{\pi\pm}$ for $\tilde{g}=\pi$, 
corresponding to $\Lambda>\Lambda_c$ ($E_{\rm C}=0.220 \hbar\omega_\perp $ and 
$E_{\rm J}=0.102 \hbar\omega_\perp$). The other parameters used here are $\lambda=1$, $h=4 \hbar \omega_{\perp}$, 
$R= 2 \times 10^{-3} (\hbar \omega_{\perp}^{2}/a_{\perp})$.}
\label{fig:2}
\end{figure}

The tunnelling behaviour of different initial states such as $\Psi_g$ and $\Psi_e$ under a potential gradient can be explained
 in terms of the time-independent solutions \cite{sak}. These can be plotted as a function of time via their dependence on the 
time-dependent potential gradient $\delta=Rt$. The time evolution and the corresponding time-independent 
population difference for state $\Psi_e$ is shown in Fig. 3(a), from which it is found that, for slow velocities, 
the system follows the eigenstate almost adiabatically. 
The initial dynamics discussed above is also well described by the two-state
 model \cite{Two_State_0,Two_State_1,Two_State_2,shen,Pendulum,Pi_BEC,Two_State_3,Two_State_4,Two_State_5,Janne,kohler,Sols_NJ}
 (grey line in Fig. 3(a)). However, for larger gradients the full potential GP calculation 
predicts that the atoms return to the lower (or left) potential well, as illustrated by the density snapshots in Fig. 3(b), 
whereas the two-state model suggests they remain in the upper (right) well. This breakdown of the two-state model occurs because 
it does not take higher lying modes into consideration \cite{sak}. 
This is an important consideration for any experimental demonstration of macroscopic flow to the higher well.

\begin{figure}[h]
\centering
\includegraphics[width=5.0cm,height=6.88cm]{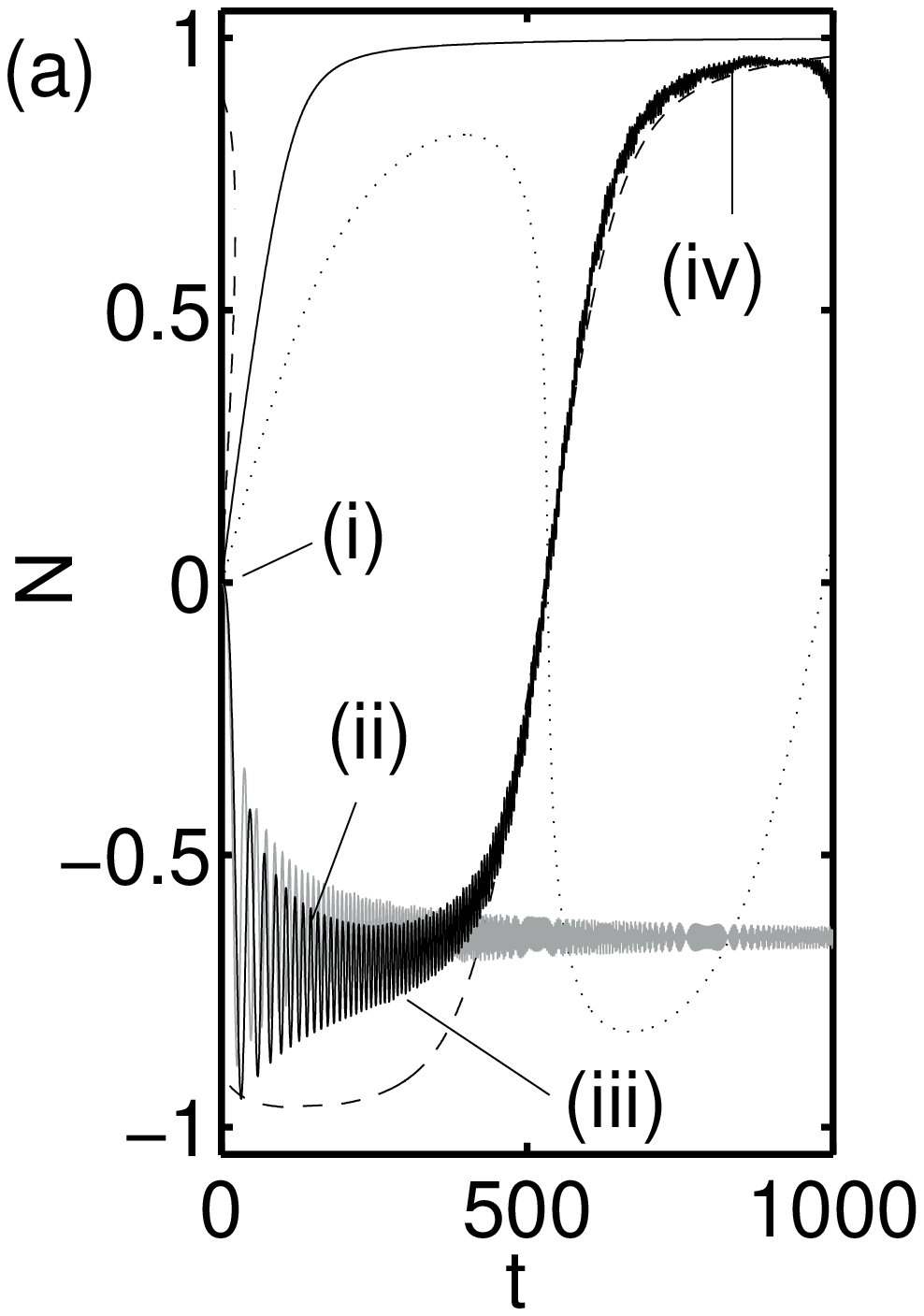}
\includegraphics[width=6.84cm]{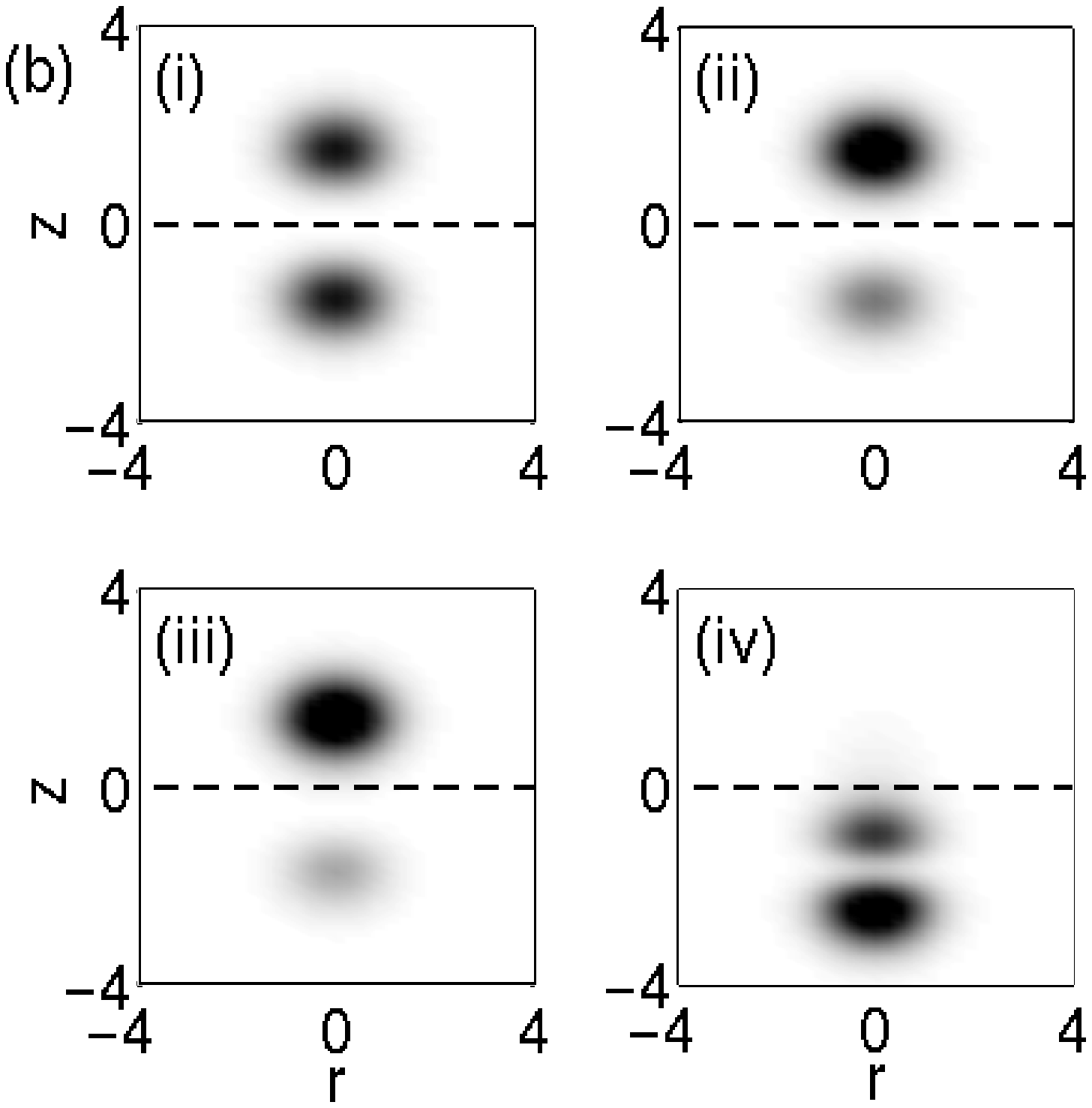}
\caption{(a) Evolution of fractional population difference $N$ as a function of time, 
for a system initially prepared in state $\Psi_{e}$ based on the Gross-Pitaevskii equation (black line) 
and the two-state model, Eq.~(\ref{eq:2state}) (grey line) with initial condition $N(0)=0$ and $\phi(0)=\pi$. 
The fractional population differences for the eigenstates are also plotted 
for the ground (solid line), first excited (dashed line) and second excited states (dotted line). 
Here  $h=4 \hbar \omega_{\perp}$, $\lambda=1$, $\tilde{g}=\pi$ and the potential gradient $\delta=Rt$ increases 
at constant rate $R=10^{-3}  (\hbar \omega_{\perp}^{2} /a_{\perp})$. 
(b) Snapshots of the evolution of the density distribution for case (a) 
when  (i) $t= 0$, (ii) $100 $, (iii) $300 $ and (iv) $800$ in units of $ (\omega_{\perp}^{-1})$. 
The population of both wells is initially equal ($t=0$). As the gradient is 
increased in (ii), (iii), population starts being transferred towards the right ($z>0$), upper well. Increasing
 the asymmetry beyond a threshold value
leads the population to be once again transferred to the left ($z<0$), lower well. Eventually, (iv), a transition
to the second excited state occurs (d) (see movie).} 
\label{fig:4}
\end{figure}

The flow towards the right (higher) potential well shown in Fig. 3(b) provides a clear 
macroscopic demonstration of quantum tunnelling. To consider whether this 
effect is observable in current experimental set-ups, we have studied the effect of varying the 
nonlinearity, trap geometry, and the time dependence of the ramp. 
Note that, the effect of interactions has also been considered in \cite{garraway}, with the 
effective interaction also modified by atom losses \cite{kohler}. Our studies reveal that increasing the nonlinearity 
causes a reduction in the amount of initial flow to the upper well, and thus tends to inhibit 
the experimental observation (see next section for experimental estimates).
For example, using the parameters of Fig. 3 with a nonlinearity ten times larger 
(i.e. $\tilde{g} =10 \pi$), leads to a reduction of the average population imbalance 
induced by the applied potential gradient of slightly more than a factor of 2. 
It is thus natural to ask if other factors (e.g. modifying 
initial trap aspect ratio, or changing barrier height $h$) will have the 
opposite effect. Enhanced tunnelling has been predicted for `pancake' traps ($\lambda > 1$) \cite{MF_3}. 
Furthermore, such geometries feature a larger energy splitting between the ground 
and first excited state, making them more 
robust to external perturbations (e.g. thermal \cite{Two_State_3,Sols_NJ,Janne,SolsBECbook}). 
However, the suppression of tunnelling induced by increased nonlinearities 
cannot be compensated by changing the geometry. We should further comment on the extent to which the above findings depend on
 the rate $R$  with which the linear potential gradient $\delta=Rt$ is applied. 
Fig. 3(a) shows the dependence for  $R=10^{-3}  (\hbar \omega_{\perp}^{2}/a_{\perp})$. 
If $R$ is increased by a factor of 10 then the  maximum flow to the right well is reduced 
by roughly a factor of 2. Also flow to higher potential region can only be observed for approximately one tenth 
of the time, unless the gradient is ramped up to a particular value and subsequently kept constant.
If the gradient is kept constant at the point of maximum population difference, the population remains trapped 
in the right upper well, i.e., macroscopic quantum self-trapping
 \cite{Two_State_1,Two_State_2,shen,elena} occurs to a state with higher potential energy. 
In this regime, where the gradient does not exceed the value at which the flow is reversed, the two-state model
 correctly predicts the behaviour.

\section{Experimental Considerations}
\label{sec:expt}

We now discuss the feasibility of observing flow to the upper well using phase imprinting \cite{Phase_Imprinting,Reinhardt}. 
Starting from the condensate ground state in a symmetric double-well trap, population can be transferred to the excited states 
by applying a light-induced potential of the form
\begin{equation}
V_{r}\left(z,t\right)=\alpha \sin \left(\pi t/\tau_{0}\right)\tanh \left(z\right)
\end{equation}
for ($t < \tau_{0}$), where $\alpha$ and  $\tau_{0}$ are constants which we vary. At $t=\tau_{0}$, the 
potential $V_{r}$ is suddenly switched off such that there is a $\pi$ phase shift between the two wells. 
This simple phase-imprinting method does not distinguish between states with similar density and phase profiles such 
as $\Psi_e$ or $\Psi_{\pi\pm}$. Other more sophisticated methods of preparing the initial state 
such as 2-photon adiabatic passage \cite{mholland} could also be explored.

We choose the phase imprinting parameters such that the amplitude of the subsequent number oscillations between 
the wells in a symmetric double-well potential (i.e. in the absence of a potential gradient) are minimized. 
The time dynamics for this case are shown by the grey lines in Fig. 4, and essentially correspond to $\pi$-oscillations
 with $\langle N(t) \rangle \neq 0$, as discussed in \cite{Pendulum,Two_State_2,shen,Pi_BEC,Exact_Phase} and shown in Fig.~2(c). 
For an imprinted phase of $\pi$, the population oscillates with most of the 
condensate in the left well (grey line in Fig. 4(a)), 
while for an imprinted phase of $-\pi$, the population oscillations are contained within 
the right well (grey line in Fig. 4(b)). In both cases, the addition of the potential gradient at a time
 indicated by the open circle in Fig.~4, induces a flow to the right or upper 
potential well (solid lines in Fig.~4). Even at the time in the $\pi$-oscillation
 cycle when most of the population is already on the right well 
and would subsequently flow back to the left, the addition of the gradient induces more flow 
to the right, as shown in Fig. 4(b). Note that, in this case, the population 
remains trapped in the right well until the influence of the second excited state becomes important. 
If the correct initial state parameters are obtained from the full GP calculation,
 then the results shown in Fig. 4 can be reproduced using the two-state model, except for the transition to the 
second excited state. However, a full potential calculation is required 
to correctly predict the initial state and the dynamics for 
larger potential gradients, when the two-state model breaks down.

\begin{figure}[hbt]
\centering
\includegraphics[width=12.0cm]{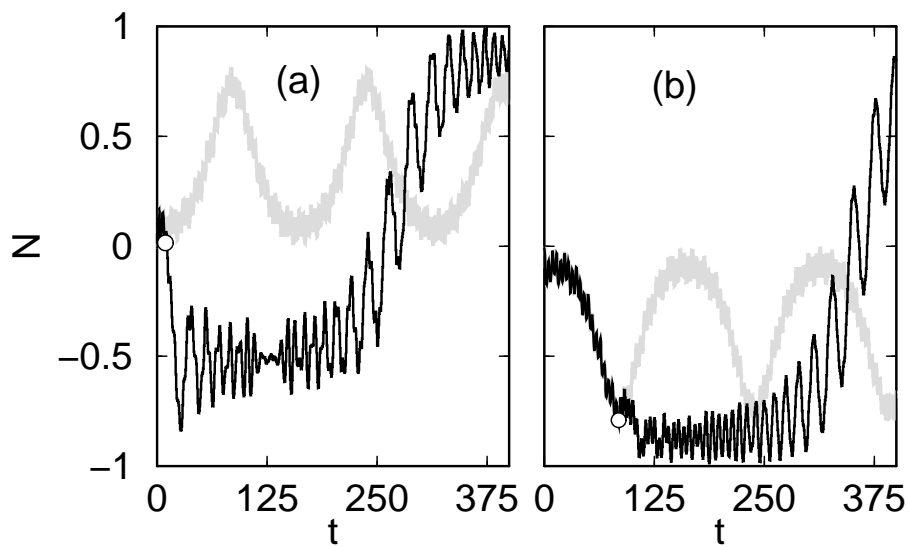}
\caption{Evolution of fractional population difference $N$ as a function of time for initial states prepared 
by phase imprinting ($|\alpha|=\hbar \omega_{\perp}$, $\tau_{0} = 3 \omega_{\perp}^{-1}$). 
The grey and black curves correspond respectively to the absence of a potential gradient (i.e. symmetric 
double well), and the addition of a potential gradient $\delta=R(t-\tau_{1})$ increased linearly at
rate $R=2 \times 10^{-3}(\hbar \omega_{\perp}^{2} /a_{\perp}) $
 from time $\tau_{1}$, with this time indicated by the open circles. 
(a) $\alpha=\hbar \omega_{\perp}$, $\tau_{1}=10 \omega_{\perp}^{-1}$ and
 (b) $\alpha=-\hbar \omega_{\perp}$ $\tau_{1}=85 \omega_{\perp}^{-1}$. Other parameters as in Fig.~2(c).}
\end{figure}

Finally, we discuss typical experimental parameters required for the demonstration of
 Josephson flow to the upper potential well. The number of atoms is given by
\begin{equation}
{\cal N}={\frac{\tilde{g}}{4\pi}}{\frac{a_{\perp}}{a}}=
{\frac{\tilde{g}}{4\pi a}}\sqrt{{\frac{\hbar}{m\omega_{\perp}}}}~.
\label{eq:numberpart}
\end{equation}
The total atom number is independent of the trap aspect ratio, therefore,
 for given dimensionless nonlinearity $\tilde{g}$, large atom numbers can be obtained 
for light, weakly-interacting, transversally weakly-confined systems. For a large number 
of atoms, one should preferably choose species with a small value of $a \sqrt{m}$. For example, taking $\tilde{g}=4 \pi$ 
and $\omega_{\perp} = 2 \pi \times 5$ Hz, we find: ${\cal N}=$  3300 ($^{23}$Na) and 1000 ($^{87}$Rb).
  An enhancement of the atom number by a factor of 10 may be possible by tuning around a Feshbach
 resonance (e.g. $^{23}$Na, $^{85}$Rb, $^{133}$Cs) \cite{Fesh}. 
In the case of $^{7}$Li and $^{85}$Rb, the number of atoms needed to observe such Josephson flow is not likely 
to exceed the critical value\cite{Rice} for collapse. 

Note that for fixed, reasonably small, nonlinearity ($\tilde{g} < 10 \pi$), such that the effect can be 
clearly observed, one needs weak transverse confinement $\omega_{\perp}$ in order to obtain a
 reasonable number of atoms which can be imaged easily. However, small  $\omega_{\perp}$ imply 
long timescales, such that the observation of this effect becomes limited 
by other factors (e.g. thermal damping \cite{Two_State_3,Sols_NJ,Janne,SolsBECbook}, atom losses \cite{kohler}, etc.). 
If we choose $\omega_{\perp} =2 \pi \times (5-100)$Hz, then the preparation 
of the $\Psi_e$ state requires a time $\tau_{0} \sim (300-150)$ms. and an applied field gradient 
$R=(10^{-3}-10^{-2})(\hbar \omega_{\perp}/a_{\perp})$ (which translates into a Zeeman shift of (1$-$100 MHz)/cm)
 ramped up over a time $t_{\rm exp} \sim$ (1.5 s$-$75 ms).

\section{Conclusions}

We have studied the Josephson dynamics of phase-imprinted condensates in a double-well potential
 in the presence of a time-dependent potential gradient. We show that phase imprinting can 
lead to a significant change in the flow direction producing a clear 
signature of macroscopic quantum tunnelling. We have discussed the range of parameters for optimum 
experimental demonstration of this effect, and find that it is only weakly-dependent on the aspect ratio of the trap. 
However, suppression of the flow for large nonlinearities, restricts the condensate size to a few thousand atoms. 
An attractive candidate for the observation of the flow reversal is the recently realized atom chips 
\cite{Atom_Chips} with a blue detuned laser beam to create the weak link. The observation of the flow of phase 
imprinted states would provide a clear experimental demonstration of the Josephson effect.

\ack We acknowledge funding from the UK EPSRC.


\end{document}